\documentclass[12pt,preprint]{aastex}
\begin{document}
\title{Using Stellar Limb-Darkening to Refine the Properties of HD~209458b}
\author{Heather A. Knutson, David Charbonneau\altaffilmark{1}, Robert W. Noyes}
\affil{Harvard-Smithsonian Center for Astrophysics, 60 Garden Street, Cambridge, MA 02138}
\altaffiltext{1}{Alfred P. Sloan Research Fellow}
\email{hknutson@cfa.harvard.edu, dcharbonneau@cfa.harvard.edu, noyes@cfa.harvard.edu}

\author{Timothy M. Brown}
\affil{Las Cumbres Observatory, 6740 Cortona Drive, Goleta, CA 93117}
\email{tbrown@lcogt.net}

\and

\author{Ronald L. Gilliland}
\affil{Space Telescope Science Institute, 3700 San Martin Drive, Baltimore, MD 21218}
\email{gillil@stsci.edu}

\begin{abstract}

We use multi-band photometry to refine estimates for the planetary radius and orbital inclination of the transiting planet system HD~209458.  We gathered 1066 spectra over four distinct transits with the STIS spectrometer on the \emph{Hubble Space Telescope} using two gratings with a resolution $R=1500$ and a combined wavelength range of $290-1030$~nm.  We divide the spectra into ten spectrophotometric bandpasses, five for each grating, of equal wavelength span within each grating, and fit a transit curve over all bandpasses simultaneously.  In our fit we use theoretical values for the stellar limb-darkening to further constrain the planetary radius.  We find that the radius of HD~209458b is  $1.320\pm0.025$~$R_{\rm Jup}$, which is a factor of two more precise than current estimates.  We also obtain improved estimates for the orbital period $P$ and time of center of transit $T_C$.  Although in principle the photon-limited precision of the STIS data should allow us to measure the timing of individual transits to a precision of $2-7$~s, we find that uncertainties in the stellar limb-darkening coefficients and residual noise in the data degrade these measurements to a typical precision of $\pm 14$~s.  Within this level of error, we find no significant variations in the timing of the eight events examined in this work.

\end{abstract}


\section{Introduction}\label{intro}

HD~209458b is the best-studied transiting planet to date, due in large part to its proximity and the resultant apparent brightness ($V=7.6$) of its parent star.  Although HD~209458b's physical characteristics have been measured more accurately than those of the other eight transiting planets \citep{char00,henry00,jha00,brown01,witt05,winn205}, it is the only transiting planet whose radius is not consistent with the predicted values for irradiated hot Jupiters.  \citet{guil96} predicted that hot Jupiters like HD~209458b, which orbit at typical separations of 0.05 AU from their parent stars, would have radii that are inflated relative to Jupiter.  The radius of HD~209458b was $20\%$ larger than predicted, however, even including the effects of increased irradiation \citep{bod01,show02,bod03,bar03,laugh05}.  In contrast, the radii of other transiting planets are within the predicted range of values for irradiated hot Jupiters \citep{tor04,mou04,pon04,kon04,laugh05,sato05,char06,bou05,bak06}.

The precise determination of the planet's radius is also of critical importance to correctly interpret the results of several follow-up observations of the planetary atmosphere \citep[e.g.][]{char06a}, notably through transmission spectroscopy \citep{char02,dem05a}, upper limits on reflected light \citep{rowe06}, and the study of the planetary thermal emission through secondary eclipse monitoring \citep{dem05,rich03a,rich03b}.
 
\subsection{Proposed Models for HD~209458b}\label{models}

Several models have been proposed to explain the source of additional internal energy required to reproduce the observed size of HD~209458b.  \citet{show02} argue that the intense radiation from the star might create strong winds in the planet's atmosphere, which would transport energy downward and heat the planet's interior to higher temperatures than would otherwise be expected.  They calculate that $1\%$ of the incident stellar radiation would have to be converted into winds to explain the observed radius of HD~209458b.  The majority of the transiting planets that have been found to date do not have radii that are significantly larger than the predicted values for irradiated hot Jupiters, however, which indicates that either (1) winds of this strength do not occur on all hot Jupiters or (2) the size of the cores in the other (normal) planets must be increased to keep their radii the same even with the additional energy input from these winds \citep{guill06}.  If the first explanation is correct, there must be some kind of distinguishing characteristic that would allow significant winds on some hot Jupiters but not others.  If the second explanation is correct, most hot Jupiters would need significantly larger solid cores than currently predicted by current formation models \citep{guill06}.

\citet{bod01,bod03} propose that the required energy may be provided by dissipation from tidal circularization of an eccentric orbit.  \citet{bod03} calculate that an average eccentricity of 0.03 would explain HD~209458b's enlarged radius if the tidal quality factor $Q\approx 2.5\times10^5$.  This is comparable to the tidal quality factor for Jupiter, which is estimated to be between $10^5$ and $10^6$ \citep{gold66,yod81}.  \citet{laugh05b} measure an eccentricity of $0.018\pm0.009$ using radial velocity data from 2004/2005, and use Monte Carlo simulations to show that this value is in fact consistent with zero.  \citet{dem05} also found no evidence for a non-zero eccentricity when they observed the secondary eclipse with the \emph{Spitzer Space Telescope}, constraining $e\sin{\omega}$ to be less than $0.0015$, where $e$ is the eccentricity and $\omega$ is the longitude of periastron.  Thus the planet cannot have a significantly non-zero eccentricity unless the semi-major axis is precisely aligned with our line of sight, and even then it is still significantly constrained by the radial velocity data.  

Although these results indicate that the planet is unlikely to have a significant constant non-zero eccentricity, it is possible that the eccentricity of the planet might be varying with time.  Because the time scale for orbital circularization is shorter than the age of the system by a factor of 100 \citep{cody02,bod03}, one way to produce a non-zero eccentricity for HD~209458b would be to have a second planet pumping up its eccentricity \citep{bod01,bod03}.  If the eccentricity of HD~209458b is in fact being pumped up by interactions with a second planet, \citet{mir02}, \citet{hol05} and \citet{agol05} show that changes in the eccentricity and other orbital parameters of the planet will produce short term oscillations in the timing of successive transits.  \citet{hol05} show that for HD~209458b the time scale for these oscillations is on the order of 10-100 orbital periods, depending on the orbital parameters of the second planet.  The radial velocity points from \citet{laugh05b} are widely spaced relative to the length of a single planetary orbit, and although they allow for a strong constraint on the average eccentricity over the five-year period covered by the data, the constraints that they can place on the eccentricity of an individual orbit are much weaker, as they would be determined by the errors on a relatively small (in most cases less than five) number of radial velocity measurements.  Similarly, the observations by \citet{dem05} only span a single secondary eclipse and could have occurred during a period of low eccentricity.  In \S\ref{timing_results}, we measure the locations of eight transits to check for variations in transit time that would indicate a time-varying eccentricity.    

\citet{winn05} propose a third possible energy source for HD~209458b.  They argue that the planet may be in a Cassini state, in which its spin precession resonates with its orbital precession.  In the particular Cassini state that the authors describe, the planet's spin axis would be tilted by close to $90\degr$ with respect to the normal of the plane of its orbit, and these two vectors would remain coplanar with the spin axis of the star as the system precessed.  Because the time scale for precession is significantly longer than the orbital period of the planet, the direction of the planet's spin axis would be effectively fixed over the course of a single orbit.  This would cause the substellar point to move from one pole to the other pole during a single orbital period.  \citet{bod03} calculate that a power of $10^{27}$ ergs s$^{-1}$ is required to explain HD~209458b's enlarged radius if the planet has a dense core and a power of $10^{26}$ ergs s$^{-1}$ if it does not.  According to \citet{winn05}, if the ratio of the tidal quality factor $Q$ to the displacement Love number $h$ is between $10^6-10^7$, a $90\degr$ obliquity would provide enough energy to explain the planet's enlarged radius.  This would be consistent with the higher end of current estimates for the tidal quality factor of Jupiter, which as we noted in a previous paragraph range from $10^5$ to $10^6$, and for the displacement Love number, which is estimated to be $\geq0.6$ \citep{gav77}.  Unfortunately, direct determination of the Cassini state of HD~209458b is beyond the capabilities of current techniques.

\subsection{Measuring the Radius of HD~209458b}\label{radius_measurement}

Since the fundamental quantity that each of these proposed models seek to explain is the planetary radius, a refined value is of great interest.  The current best estimates come from \citet{witt05} and \citet{winn205}.  \citet{witt05} use a combination of radial velocity data (51 published and unpublished values) and transit curves (\emph{Hubble Space Telescope} STIS and FGS data, as well as data from several ground-based observatories), which they fit simultaneously to obtain a radius of $1.35\pm0.07$~$R_{\rm Jup}$\footnote{All measurements in this paper use Jupiter's equatorial radius at 1 bar, 71492 km}.  \citet{winn205} also fit radial velocity and transit data simultaneously, but use 85 radial velocity measurements from \citet{laugh05} and transit data from the \emph{Hubble Space Telescope} STIS observations by \citet{brown01} alone.  The authors also include the constraints on the timing of the secondary transit from \citet{dem05}.  Fitting these three data sets simultaneously, they find that the radius of HD~209458b is $1.35\pm0.06$~$R_{\rm Jup}$.  The results of \citet{witt05} and \citet{winn205} are virtually identical, despite differences in the data used in the fit.  This is because the radius of the planet is determined by fitting the photometric transit curves (radial velocity data decreases the uncertainty in other parameters, and thus has a marginal effect on the best-fit value for the radius of the planet, but this effect is small).  The high quality of the \citet{brown01} \emph{HST} data, which both authors use, relative to data from ground-based telescopes means that the \emph{HST} data dominate any simultaneous fit of available transit curves, explaining the similarity of these results.

In this work we improve on previous measurements of the planet's radius by fitting a new photometric data set, gathering 1066 spectra over four distinct transits with the STIS spectrometer on the \emph{Hubble Space Telescope}.  The increased wavelength range of our data \citep[we use two gratings with a combined range in wavelength space of $290-1030$~nm, instead of the single grating spanning $581.3-638.2$~nm used by][]{brown01} means that we can divide the spectra into ten spectrophotometric bandpasses and bin to create ten individual photometric timeseries, while still maintaining a photometric precision comparable to the \citet{brown01} data.  The relative RMS variation for the out-of-transit data in our ten bandpasses ranges from $1.5-5.8\times10^{-4}$ for a typical cadence of 40~s, with the majority of bandpasses closer to the lower value.  This noise is only slightly higher than $1.1\times10^{-4}$ for a cadence of 80~s, the typical value reported by \citet{brown01}.  

Unlike \citet{brown01}, who fit data spanning a single bandpass, we fit a transit curve over ten bandpasses simultaneously.  This allows us to break a fundamental degeneracy in the shape of the transit curve.  As \citet{jha00} observed, the problem arises from the fact that a transit is primarily described by its duration and depth.  For observations in a single bandpass, it is possible to fit the same transit curve with a larger planet, simply by increasing the stellar radius and decreasing the inclination (the converse is true for smaller planetary radii).  By observing the same transit in multiple bandpasses, we are able to determine the inclination uniquely, independent of assumptions about the stellar and planetary radii.  

\citet{jha00} use a similar approach in their analysis of multi-color $BVRIZ$ photometric data from ground-based telescopes, obtaining a value of $1.55\pm0.10$~$R_{\rm Jup}$ for the radius of HD~209458b.  \citet{deeg01} also fit data from ground-based telescopes spanning the four Str\o mgren bandpasses, and find a value of $1.435\pm0.05$~$R_{\rm Jup}$ for the planet's radius.  Although both authors obtain a value for the radius of the planet from their fit, they differ in their initial assumptions.  The goal of \citet{jha00} was to obtain an improved value for the radius of the planet, and so they use stellar models to predict the shape of the limb-darkened light curves rather than fitting for the limb-darkening coefficients.  In contrast, the primary goal of \citet{deeg01} was to use the transit curve to probe the limb darkening of the primary star, and so they used a linear limb-darkening law and left the limb-darkening coefficients as free variables in their fit.  In this work we use a similar approach to that of \citet{jha00} and calculate four-parameter non-linear limb-darkening coefficients from theoretical models.  We examine the accuracy of our limb-darkening model in more depth in \S\ref{transit_curve_fit}. 

Although we can measure the inclination independently, there is still an additional degeneracy between the radius of the planet ($R_P$), the radius of the star ($R_{\star}$), and the mass of the star ($M_{\star}$), where $R_P \propto R_{\star}\propto M_{\star}^{1/3}$.  The depth of the transit curve is a function of $R_P/R_{\star}$, so the only piece of information needed to break this degeneracy is the mass of the star.  This can be determined by comparing the observational parameters of HD~209458, including its spectrum and absolute visual magnitude, with theoretical stellar-evolutionary models.  However, \citet{cody02} point out that the line of degeneracy between the stellar mass and radius from models at a constant luminosity is nearly orthogonal to the constraint on the stellar mass and radius from a fit of the light curves.  By combining information from the models with the constraint from the transit curves themselves, we can reduce the uncertainty in the final measurement of the mass and radius of the star, which reduces the formal uncertainties in our measurement of the planet's radius.

This is not an entirely new approach; early ground-based studies of HD~209458 assumed values for the mass and radius of the star calculated from models, and fit for the radius of the planet and inclination of its orbit alone \citep{char00,henry00}. As higher-quality data from the \emph{Hubble Space Telescope} became available, \citet{brown01}, \citet{winn205} and \citet{witt05} assume a value for the mass of the star and use this data \citep[originally published by][]{brown01} to measure the value for the stellar radius directly from the transit.  The quality of the data is important because it is the shape of the ingress and egress that breaks the degeneracy between inclination and stellar radius for data in a single bandpass.  Although our data allow us to fit for the stellar radius directly (given a value for the stellar mass), with an accuracy greater than that of \citet{brown01}, in our analysis we incorporate the additional joint constraint on the stellar mass and radius from models \citep{cody02} in order to minimize the uncertainty in the measurement of the planetary radius.  

\section{Observations and Data Reduction}\label{obs_red}
 
\begin{figure}
\epsscale{0.8}
\plotone{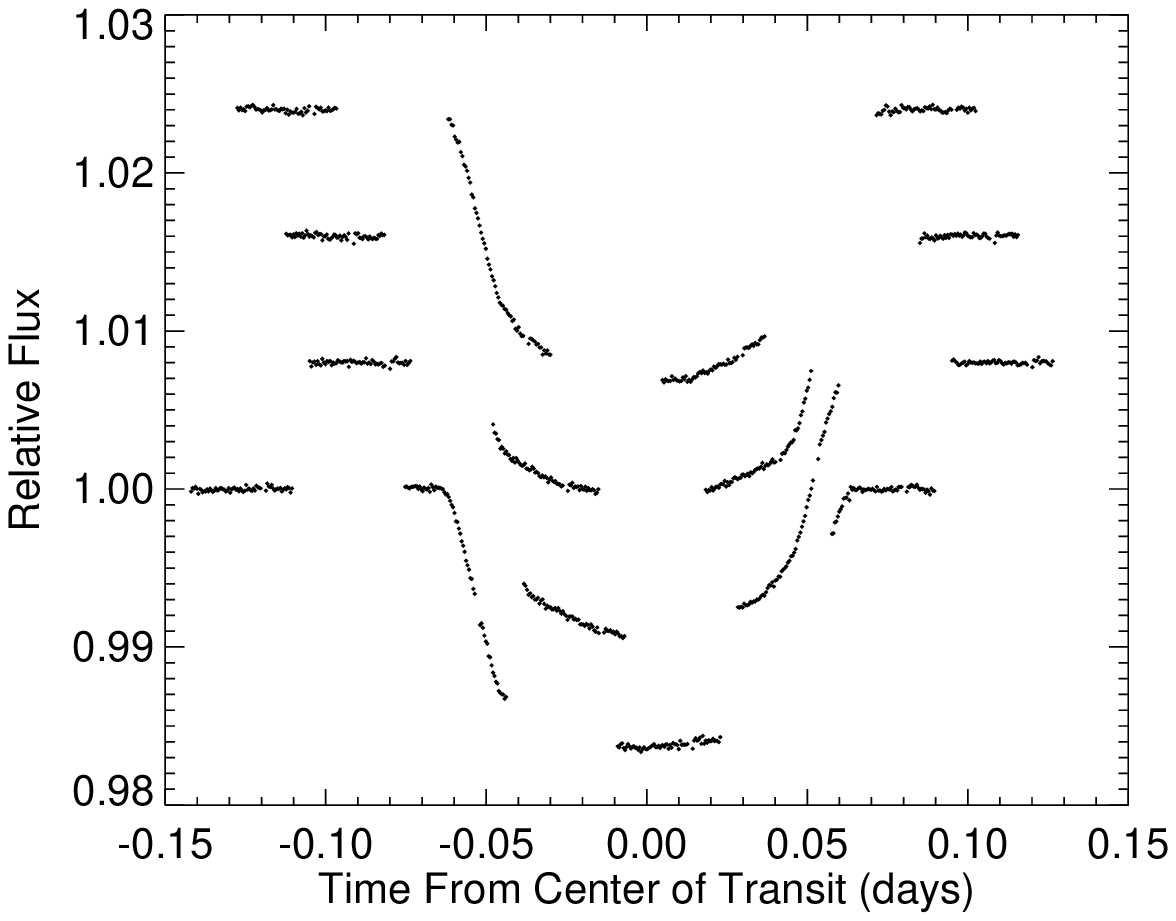}
\caption{Normalized timeseries spanning the entire wavelength range of the grating for the four transits observed.  Transits occurred (from top to bottom) on UT 2003 May 3 (HJD 2452763.7, G430L grating), May 31 (HJD 2452791.4, G750L grating), June 25 (HJD 2452816.1, G430L grating), and July 5 (HJD 2452826.6, G750L grating).  Each successive transit is offset by 0.008 in relative flux.\label{unbinned_data}}
\end{figure}

\begin{figure}
\epsscale{0.8}
\plotone{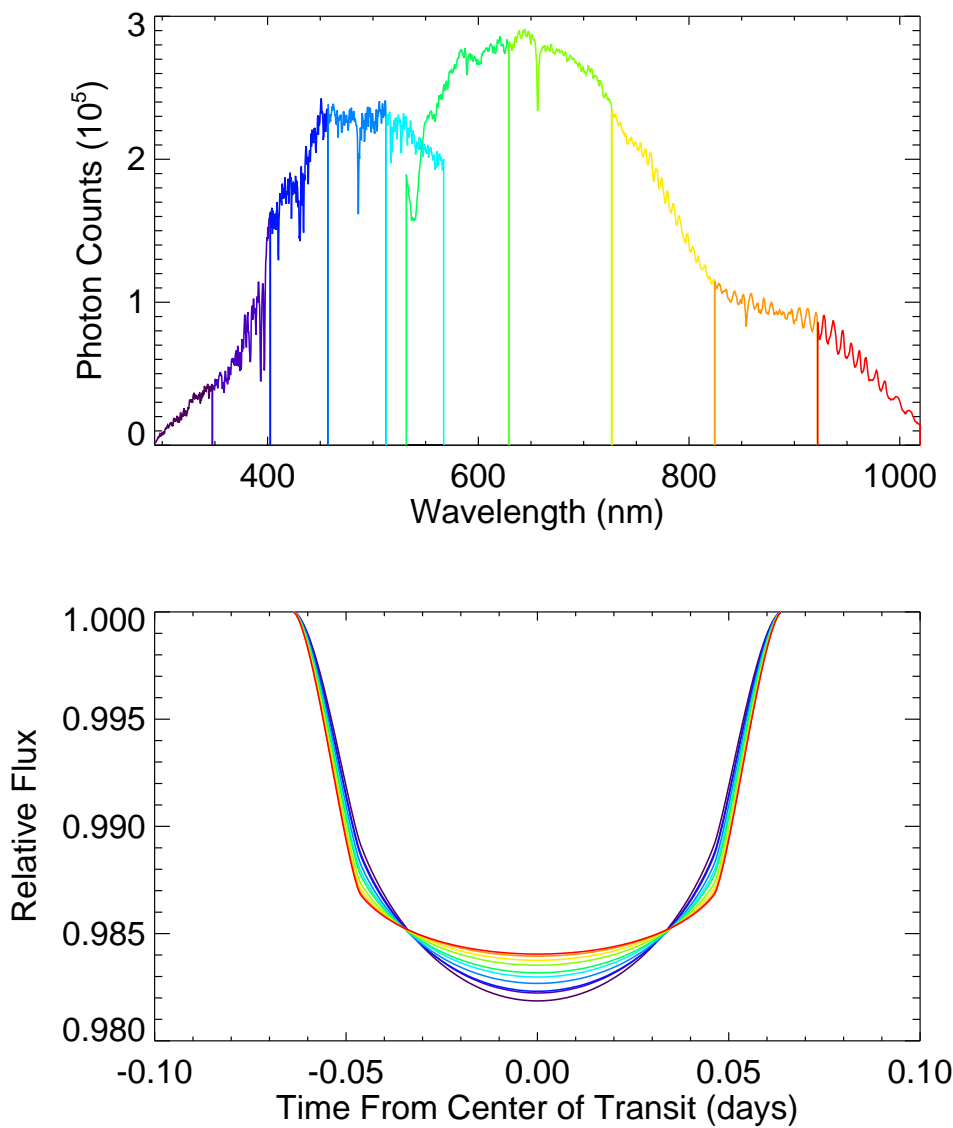}
\caption{TOP: Two typical extracted one-dimensional spectra.  The G430L spectrum spans the wavelength range from $293-567$~nm, and the G750L spectrum spans the wavelength range from $532-1019$~nm.  The vertical lines denote individual bandpasses within each grating, with some overlap between gratings around 550~nm.  Fringing from the detector is significant for wavelengths longer than 750~nm.  BOTTOM: Limb-darkened light curves for the ten bandpasses described above, as discussed in \S\ref{transit_curve_fit}, plotted using the best-fit parameters from Table \ref{best_fit_coeff} and model four-parameter non-linear limb-darkening coefficients.\label{spectrum_limbdk}}
\end{figure}

\begin{figure}
\epsscale{0.8}
\plotone{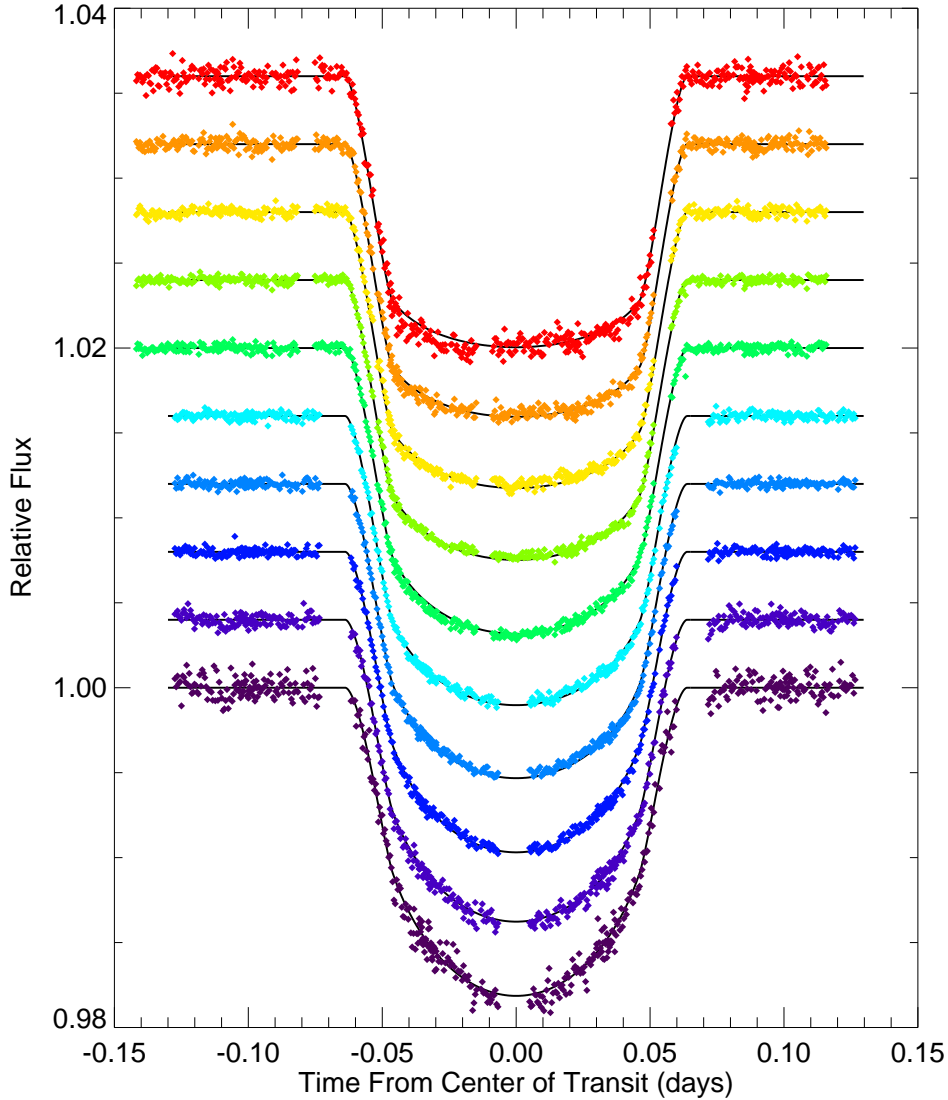}
\caption{Normalized data for the ten bandpasses shown in Figure \ref{spectrum_limbdk}, with theoretical transit curves using the best-fit parameters from a simultaneous fit of all bandpasses (Tables \ref{best_fit_coeff} and \ref{ephemeris}) and model four-parameter non-linear limb-darkening coefficients overplotted.  Note that each bandpass contains data from two separate visits, consisting of four spacecraft orbits each (visits are plotted individually in Figure \ref{unbinned_data} for reference).  Each successive transit curve is offset by 0.004.  Although the transit curves are a good fit for most of the data, there are some systematic deviations on timescales comparable to that of a \emph{HST} orbit (see Figure \ref{transit_residuals}) that are discussed in \S\ref{timing_results}.\label{binned_data}}
\end{figure}

We obtained twenty \emph{HST} orbits of STIS spectra, grouped into four ``visits'' of five consecutive orbits each, during which the telescope was pointed continuously at the star.  Each visit was timed so as to span one complete transit of HD~209458b.  These observations were obtained during the period from UT 2003 May 3 to July 6 (see Figure \ref{unbinned_data}) as part of the GO-9447 program.  For a more detailed discussion of the suitability and performance of the STIS spectrometer for transit photometry see \citet{brown01}.  We observed using both the G430L and G750L gratings, with two visits in each grating.  Together these two filters cover a combined range of $290-1030$~nm, with some overlap between filters around 530~nm (see Figure \ref{spectrum_limbdk}).  In our analysis we sub-divide the spectrum and bin it into five equally spaced bandpasses in wavelength space within each grating, yielding a total of ten distinct bandpasses (see Figure \ref{binned_data}).  We note that there is significant fringing at wavelengths longer than 800~nm from the internal interference of the thin STIS CCD, but we find this fringing has a negligible effect on the binned data.  We gathered spectra with a 22~s integration time at a cadence of 42~s in the G430L grating and a 19~s integration time and 39~s cadence in the G750L grating within the part of each spacecraft orbit where the star was visible, yielding excellent time resolution during the crucial ingress and egress periods.

We assigned wavelengths to our data by cross-correlating our spectra (using the wavelength solution returned by the automated STIS pipeline) with the model spectrum of HD~209458 described in \S\ref{transit_curve_fit}.  We performed this analysis over 20 individual bandpasses spanning the spectral range of the data.  We found that the typical offset did not vary by more than one pixel in time or with spectral region, and hence we calculated the mean offset value and shifted the STIS wavelength solution accordingly.

We create our timeseries for each bandpass by binning the portion of each individual $1024\times64$ pixel spectrum within the desired range of wavelengths to create a single photometric measurement.  We optimized the size of our bin in the cross-dispersion direction to minimize the contribution from read noise and scattered light distributed across the detector array.  This was particularly important in the regions of the spectrum where the signal was small, as the additional signal obtained from a larger bin in the cross-dispersion direction was minimal compared to the increase in noise.  

We optimized our aperture sizes in the cross-dispersion direction for 50-pixel-wide sections spanning the range of the spectrum.  In order to determine the optimal size, we picked an aperture, binned the data to create a timeseries, and measured the relative variation in the out-of-transit data for that timeseries.  We repeated this process systematically for all possible aperture sizes, ranging from $3-64$ pixels, and selected the aperture that minimized this variation.  The optimal aperture sizes in the cross-dispersion direction for each of our segments ranged from $5-31$ pixels, with a median size of 19~pixels.  The narrowest apertures were in the low-flux regions at the edges of the spectra and the wider apertures in the central high-flux regions.

 Once we had determined our optimal apertures and binned our spectra within each bandpass to create a photometric timeseries, we found that there were several prominent trends remaining in the data. First, the flux values increased gradually over time during each visit, so that the average flux before the beginning of a typical transit was 0.1\% lower than the average flux after the transit.  This trend was particularly pronounced in data taken during the first spacecraft orbit of each visit, which was sometimes as much as 0.3\% lower than data at the end of the visit.  We attributed this trend to thermal settling of the telescope at its new pointing.  Second, the first point of each spacecraft orbit was typically 0.2\% lower than subsequent points.  Third, the data also showed a trend within individual spacecraft orbits, with a typical range of $0.1-0.2$\% between minimum and maximum values, that was repeated consistently from one orbit to the next.  Although it is not clear what caused these variations, they were found to correlate with spacecraft orbital phase.  This indicates that they were caused by changes in the instrument and not intrinsic variations in the star.  

In order to remove these trends, we used two steps.  First, we chose to discard the first of the five orbits in each visit and the first point in each of the remaining orbits.  The first orbit and initial points of subsequent orbits consistently exhibited the largest systematic effects, and discarding these points still left sufficient data before and after the transit to measure an accurate value for the out-of-transit flux from the star.  \citet{brown01} removed the same points from their data, which showed similar trends.  In the second step, we removed the remaining trends by fitting the out of transit points in each transit simultaneously with a third-order polynomial function of spacecraft orbital phase and a linear function of time, and dividing the data by the resultant function.  This fit was done individually for each of the four visits, in each bandpass for that visit.

The normalized timeseries for each transit (Table \ref{data_table}) is quite flat, although as we will discuss in \S\ref{timing_results}, there are still some remaining systematic effects visible in the combined plots (the plot for each bandpass includes data from two transits overplotted) in Figure \ref{binned_data}.  The RMS variation in relative flux for the out-of-transit data in each of these bandpasses ranges from $1.5-5.8\times10^{-4}$, with the noisiest bandpasses at the low-flux ends of the spectrum.  The time coverage of the data for the combined transit curve is quite good, as shown in Figure \ref{unbinned_data}.

When we compared the RMS scatter of the out of transit data to the expected photon noise for the data in each bandpass, we found that the RMS scatter was on average 21\% higher than the photon noise.  We examined several potential noise sources.  The read noise for the STIS detector is 7.75~e$^-$~pix$^{-1}$ RMS at a gain of four, with 3800~pixels in an average bandpass.  A typical bandpass has a flux of $10^8$~e$^{-}$ and a corresponding Poisson noise of $10^4$~e$^-$ RMS, so read noise would only decrease the SNR by 0.1\%.  There is also a constant background contribution from a combination of light reflected from the Earth, zodiacal light, and geocoronal emission.  This background contributes up to 0.06~e$^{-}$~s$^{-1}$~pix$^{-1}$, for a total contribution of $10^4$~e$^{-}$.  This is only $10^{-4}$ of the total signal, so its contribution to the Poisson noise is negligible.  Although the magnitude of this background may vary over time as the position of the telescope changes relative to the various sources, this is unlikely to be the source of the additional variation, for two reasons.  First, although some components of the systematic variations in the data correlate well with changes in the width and location of the point spread function of the spectrum in the cross-dispersion direction, they do not correlate with the relative positions of the Sun and moon, two likely sources of scattered light within the instrument.  Second, when the data is binned using increasingly narrow slits in the cross-dispersion direction, it is clear that most of the systematic variations are contributed by the central 20 pixels around the peak of the point spread function, making scattered light in the instrument an unlikely explanation.  The most plausible origin is the change in the focus of the instrument and location of the source in the slit, which do not precisely repeat from orbit to orbit.  In order to properly account for all noise sources in our error estimates for the points in our timeseries, we chose to set the error for all points within a given bandpass as the standard deviation of the out of transit points for the data in that bandpass.

\section{Fitting the Transit Curve}\label{transit_curve_fit}

We use the complete analytic formula given in \citet{mand02}, without approximations, to calculate our transit curves.  The expression given in \citet{mand02} is a function of six dimensionless parameters, including the ratio of the planetary radius to the stellar radius, the impact parameter in units of stellar radii, and four nonlinear limb-darkening coefficients.  We would like to fit for the first two variables, as these variables determine the best-fit values for the mass and radius of the star, the radius of the planet, and the inclination of the planet's orbit relative to the observer.  We have designed our fitting routine to use the latter four parameters as its input, calculating the two dimensionless parameters for the analytic expression from \citet{mand02}.

Rather than fitting for the period $P$ and initial transit time $T_C$ simultaneously, we fix these parameters to an initial estimate, taken from \citet{brown01} and \citet{char03}, and fit for the best planet radius, inclination, stellar mass, and stellar radius.  We then go on to find $P$ and $T_C$ using the method described in \S\ref{ephemeris_fit}.  We then take the new best-fit $P$ and $T_C$ and repeat our fits for planet radius, inclination, stellar mass, and stellar radius.  We iterate this process until our best-fit values for all six parameters converge to a consistent solution.  Because we are using a relatively simple fitting routine (downhill simplex) to avoid the need for derivatives of our transit function, this iterative fitting ensures that the function converges to the true global minimum.  We assume the orbital eccentricity is zero, for the reasons discussed in \S\ref{intro}.  

As discussed in the introduction, we must make an initial assumption about either the mass or the radius of the star in order to break the degeneracy between these parameters and the radius of the planet when we fit our transit curve.  We constrain the mass and radius of the star based on \citet{cody02}, where $R_{pred}$ is the radius predicted by the theoretical mass-radius relation discussed in \S\ref{intro} for a given stellar mass:

\begin{equation}\label{r_star_constraint}
\frac{R_{pred.}}{R_\sun}=1.18-\frac{1}{0.96}\left(\frac{M_{\star}}{M_\sun}-1.06\right)
\end{equation}
We implement this constraint by adding two terms to the chi-squared function where we treat $M_{\star}$ and $R_{\star}$ as Gaussian random variables with standard deviations equal to the published errors.  The inclination and planetary radius have no such constraints.  The goodness-of-fit parameter is given by:
\begin{equation}\label{cq}
\chi^2=\sum_{i=1}^N\left(\frac{p_i-m_i}{\sigma_m}\right)^2+\left(\frac{\frac{M_{\star}}{M_\sun}-1.06}{0.13}\right)^2+\left(\frac{\frac{R_{\star}}{R_\sun}-\frac{R_{pred.}}{R_\sun}}{0.055}\right)^2
\end{equation}
where $m_i$ is the $\textrm{i}^{th}$ measured value for the flux from the star (with the out-of-transit points normalized to one), $p_i$ is the predicted value for the flux from the theoretical transit curve, and $\sigma_m$ is the error on each individual flux measurement, which we take to be the standard deviation of the out of transit points for that bandpass, as discussed in \S\ref{obs_red}.  $M_{\star}$ and $R_{\star}$ are the fitted values for the mass and radius of the star, and we constrain the mass of the star using the value from \citet{cody02}, $1.06\pm0.13$~$M_\sun$.  The error on the predicted radius of the star, $\pm0.055$~$R_\sun$, also comes from \citet{cody02}.  

When calculating our transit curves, we use the nonlinear limb-darkening law defined in \citet{clar00}:
\begin{equation}\label{ld_law}
I(\mu)=1-\sum_{n=1}^4c_n(1-\mu^{n/2})  
\end{equation}
where
\begin{equation}
\mu=\cos{\theta}
\end{equation}
In this case, $\theta$ is defined as the angle between the normal from the surface of the star and the direction of the observer.  We derive our four-parameter nonlinear limb-darkening coefficients from theoretical models calculated by R. Kurucz\footnote{Available at http://kurucz.harvard.edu/stars/hd209458} (personal correspondence, August 2005) using an effective temperature of 6100~K, $log(g)=4.38$, and $[M/H]=+0.014$ for the star \citep{val05}.  The model spectra used in these calculations include both continuum processes and line absorption \citep{kur05}.  The limb-darkening coefficients for each bandpass are determined by taking the weighted average of the limb-darkening coefficients at individual wavelengths, with the weights determined by the measured flux at that wavelength (see Table \ref{limb_dk_coeff} and Figure \ref{spectrum_limbdk}).  This includes the wavelength-dependent response function of the STIS instrument.  As expected, we find that our limb-darkened curves are similar to those of the Sun at longer wavelengths, but exhibit greater center-to-limb variation at short wavelengths where the relative difference in intensity between blackbody curves at the temperature of the two stars is greatest (see Table \ref{solar_vs_model} and Figure \ref{binned_data}).  

To determine the best-fit radius for the planet, we evaluate the $\chi^2$ function over all ten bandpasses simultaneously, using the same values for the planetary radius, stellar mass and radius, and inclination.  Minimizing this function, we find that the best-fit value for the radius of the planet is  $1.320\pm0.025$~$R_{\rm Jup}$, with an orbital inclination of $86{\fdg}929\pm0{\fdg}010$.  The best-fit mass and radius for the star are $1.101\pm0.066$~$M_\sun$ and $1.125\pm0.023$~$R_\sun$, respectively.  The reduced $\chi^2$ for this minimized function is 1.28.

\begin{figure}
\epsscale{0.8}
\plotone{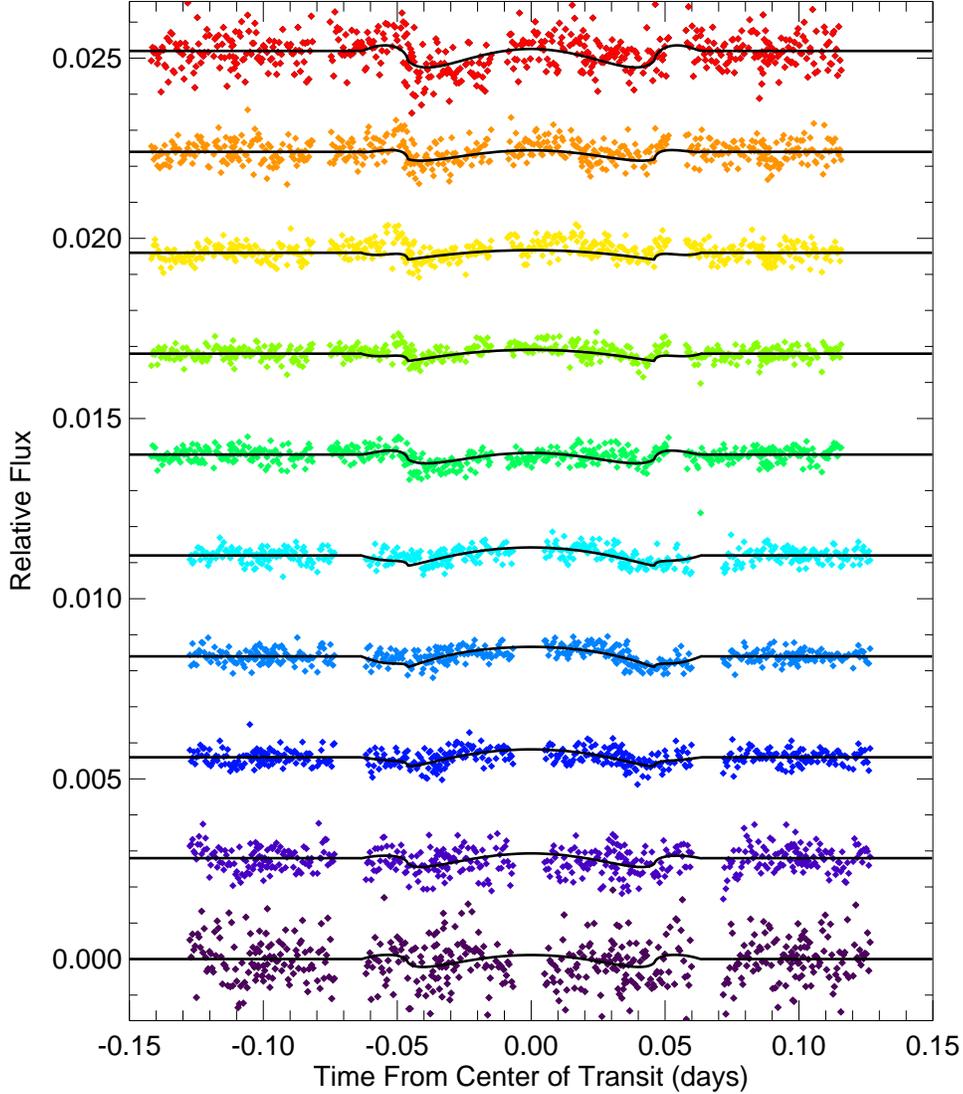}
\caption{These are the residuals in each bandpass for our best-fit parameters.  Each successive bandpass is offset by 0.0028.  As noted in \S\ref{timing_results}, the relative fluxes shift by $10^{-4}$ on average from one orbit to the next.  Each continuous section of data consists of two spacecraft orbits from different visits.  The five redmost (upper) curves were all gathered simultaneously, resulting in the correlated variations among the five bandpasses.  Similarly, the five bluemost (lower) curves were gathered simultaneously.  The solid curves overplotted are the difference between transit curves calculated using fitted quadratic limb-darkening coefficients and transit curves using the nonlinear limb-darkening coefficients derived from the model by Kurucz.\label{transit_residuals}}
\end{figure}

As a check, we also fit for the limb-darkening coefficients directly using the data.  Because we would like to compare the shapes of these curves to the curves derived using our model limb-darkening coefficients, we set the stellar mass equal to its best-fit value from Table \ref{best_fit_coeff} and continued to include an additional term in the $\chi^2$ function for the mass-radius relation from \citet{cody02}.  We also chose to fit for the limb-darkening coefficients using a quadratic limb-darkening law \citet{mand02} instead of the four-parameter non-linear law we used earlier, as the four-parameter coefficients are degenerate at the precision of our data.  In order to fit for all 24 variables, including 20 limb darkening coefficients, reliably we use a Markov Chain Monte Carlo code developed following the methods described by \citet{ford05} and \citet{teg04}, and check our results using our original $\chi^2$ minimization routine.  When we include the uncertainty in the stellar mass from Table \ref{best_fit_coeff}, we find $R_\star=1.137\pm0.024$, $R_P=1.342\pm0.032$, and $i=86{\fdg}75\pm0{\fdg}14$.  As expected, this new fit is able to remove some of the trends in the residuals, with the largest remaining discrepancies around the periods of ingress and egress (see Figure \ref{transit_residuals}).  This is not surprising, as the uncertainties in the limb-darkening laws predicted by models are greatest at the edges of the star.  

Although these new values are a better fit to the data, the uncertainties in the best-fit parameters are also correspondingly larger.  This new fit is also more sensitive to residual noise correlated with spacecraft orbital phase (discussed at the end of \S\ref{obs_red}), which can alter the shape of the ingress and egress.  As a result, we prefer the more robust fit with model-derived limb-darkening coefficients for subsequent analysis.  In an independent study, \citet{tin05} also compared the observed limb-darkening in published transits for HD~209458 to the predicted limb-darkening from the ATLAS models, and concluded that although there were some discrepancies between the two curves these discrepancies had a negligible effect on the measurement of the system parameters.  We find that in this case our values for the radius of the planet differ by less than 1$\sigma$ between the two methods.

\section{Fitting the Ephemeris}\label{ephemeris_fit}

In order to measure an accurate value for the planetary ephemeris, we included four normalized transit curves\footnote{Available at http://cfa-www.harvard.edu/$\sim$dcharbon/frames.html}  derived from STIS data taken between UT 2000 April 25 and May 13 in our analysis.  This data was taken using a higher-resolution grating (G750M instead of G430L and G750L) and spans the wavelength range from $581.3-638.2$~nm, instead of the $290-1030$~nm spanned by our data.  \citet{brown01} binned this data in a single bandpass over the entire available wavelength range.  As discussed in \S\ref{radius_measurement}, the precision of the \citet{brown01} transit curves and our transit curves are comparable.  The authors observe the same kinds of systematic effects in their data as we do in ours, although the orbit-to-orbit residuals in our data are larger, and they normalize their data with a linear function of time from first observation and a fourth-order polynomial function of orbital phase to remove these effects.  We use these normalized transit curves, taken directly from \citet{brown01}, in our fits.  

Rather than using the quadratic limb-darkening coefficients published by \citet{brown01}, we derive new four-parameter nonlinear limb-darkening coefficients for the \citet{brown01} data using the same method as described in \S\ref{transit_curve_fit}.  As before, we define our limb-darkening coefficients for the bandpass as the weighted average of the  nonlinear limb-darkening coefficients at each wavelength, with the flux from a typical spectrum at that wavelength as the weight. Although all four transits were observed with the same grating, there was a database error in the location of the subarray for the first transit, and as a result the red part of the spectrum was not entirely contained on the subarray.  We account for this misalignment by calculating one set of limb-darkening coefficients for the first visit and a different set for the other three visits.   For the first visit our four-parameter non-linear limb-darkening coefficients are $c_1=0.5838$, $c_2=-0.3430$, $c_3=0.9183$, and $c_4=-0.4240$.  For the subsequent three visits $c_1=0.5845$, $c_2=-0.3443$, $c_3=0.9179$, and $c_4=-0.4236$.  As before, we check the standard linear wavelength solution from the \emph{HST} pipeline by comparing these spectra to model spectra for HD~209458.  Using the sodium lines at 588.995~nm and 589.592~nm as our calibration we find the \emph{HST} wavelength solutions for this line are shifted by 1~pixel (0.06~nm) relative to the rest wavelength of the line, and correct the wavelength solutions accordingly before calculating our limb-darkening coefficients.

We fit for the locations of each of the eight transit centers individually using a standard $\chi^2$ function and the best-fit values for planetary radius, inclination, and stellar mass and radius from \S\ref{transit_curve_fit}.  We bin the data from 2000 into a single bandpass for each transit, and leave the data from 2003 in the same bandpasses used in the previous analysis, fitting simultaneously over all bandpasses.  After deriving the best-fit locations for each transit center, we fit for the period and initial transit location $T_C$ by plotting the number of transits versus the location of each transit in HJD and fitting the points with a line.  The constant coefficient of this linear fit gives us the best-fit $T_C$, while the slope of the line gives the period.  We then repeat our earlier fits for $R_P$, $i$, $R_{\star}$, and $M_{\star}$ using this new value for the period and initial transit time, iterating until all values converge to a consistent result. 

\section{Discussion}

\subsection{System Parameters}\label{transit_curve_results}

The best-fit values for the planet's radius and orbital inclination are listed in Table \ref{best_fit_coeff}.  Figure \ref{binned_data} shows the data in each bandpass with best-fit transit curves overplotted, and Figure \ref{transit_residuals} shows the residuals from the fit using the theoretical four-parameter non-linear limb-darkening coefficients.  The new value for $R_P$, $1.320\pm0.025$~$R_{\rm Jup}$, has uncertainties that are less than half that quoted in \citet{witt05} and \citet{winn205}.  Although our radius is slightly smaller than the values given by these two authors, it is entirely consistent within their uncertainties. Using the mass given in \citet{laugh05b}, $0.69\pm0.06$~$M_{\rm Jup}$, and scaling it appropriately to account for the different value for the stellar mass from our fits, we calculate $M_P=0.64\pm0.06$~$M_{\rm Jup}$.  This gives a density for the planet of $0.26\pm0.04$~$\rho_{\rm Jup}$ or $0.345\pm0.05$~g/cm$^3$.  

We find that the inclination of HD~209458b's orbit is $86{\fdg}929\pm0{\fdg}010$.  Although our inclination is outside the errors given by \citet{winn205} \citep[there are no independent errors for the inclination given by][]{witt05}, \citet{winn205} note that the inclination depends on their particular choice of limb-darkening law.  In this case, we find that we measure a very small error for the inclination when we assume the limb-darkening coefficients are known from models.  Unfortunately, much of this improvement is lost when we allow the limb-darkening coefficients to vary in the fit.  In this case, our use of ten individual bandpasses is no longer an advantage, as limb-darkening adds two degrees of freedom to our fit for every bandpass we include.  This makes it possible to fit the data with a significantly larger range of inclinations.  It also makes the fit more sensitive to correlated noise in the data, as discussed at the end of \S\ref{obs_red}.  For this reason, we prefer the more robust fit using model limb-darkening coefficients, which provide a reasonably good fit to the data and do not have a significant effect on the value we measure for the radius of the planet.

We note that the value for the radius of the planet that we obtain from our fits directly depends on our initial assumptions about the mass and radius of the star.  Although we use the same initial estimate for the mass of the planet as \citet{winn205} and \citet{witt05}, we include an additional restriction on the stellar radius as a function of the stellar mass \citep{cody02}.  We find that best-fit values for the mass and radius of the star from our fit are $1.101\pm0.066$~$M_\sun$ and $1.125\pm0.023$~$R_\sun$, respectively.  These values differ from the best-fit values cited in \citet{cody02} that we used in our initial constraints because the transit curve provides additional information about the stellar radius as a function of stellar mass, as discussed in \S\ref{radius_measurement}.  These values are within the quoted errors for the measurements from \citet{cody02}, and are also virtually identical to the most recent independent measurements of the mass and radius of HD~209458 from \citet{val05}, who find a stellar mass and radius of $1.11\pm0.18$~$M_\sun$ and $1.122\pm0.055$~$R_\sun$.  

We also repeat our fits without any constraints on the stellar radius.  In this case, the best-fit stellar radius is determined by the value we assume for the mass of the star.  Using the relatively small mass from \citet{cody02}, we find that $R_{\star}=1.11\pm0.04$~$R_\sun$, a value that is smaller than the one measured by \citet{cody02} but still within 2$\sigma$ of that measurement.  The radius of the planet in this fit is $1.31\pm0.05$~$R_{\rm Jup}$.  The increased errors indicate that it is the joint constraint on the mass and radius of the star which produced the most signification reduction in the uncertainty in the planet's radius in our earlier fits.   When we use the mass from \citet{val05}, we find $R_{\star}=1.13\pm0.06$~$R_\sun$, and $R_P=1.33\pm0.07$~$R_{\rm Jup}$.

In our analysis so far we have assumed that there was a single universal value for $R_P$, $R_{\star}$, $M_{\star}$, and $i$.  However, $R_P$ could vary from bandpass to bandpass, as the presence of absorption lines in the planet's atmosphere might increase the depth of the eclipse in particular bandpasses.  This is particularly relevant for the long-wavelength bands in the red grating, where model atmospheres for HD~209458b predict some of the strongest absorption bands from water vapor \citep{brown01b}.  We check the validity of our previous assumption by fitting each bandpass individually.  

Our initial results were surprising; we measured a variation in the best-fit values for $R_P$ on the order of $\pm0.02$~$R_{\rm Jup}$.  Although it is tempting to dismiss this variation as within the 1$\sigma$ errors for $R_P$, it is important to remember that these errors are dominated by the error in the values of $M_{\star}$ and $R_{\star}$, quantities that do not vary from bandpass to bandpass.  Because we applied the same constraints in each of our fits over individual bandpasses, we would expect the $relative$ value of $R_P$ within each bandpass to remain consistent over much smaller scales than the overall error for the measurement.  

Closer examination of the best-fit values for the other variables in the fit revealed that each value of $R_P$ was associated with a different value for the inclination.  Of course, inclination cannot vary with bandpass, and so we repeated the fits while fixing the inclination to the best-fit value from the simultaneous fit of all ten bandpasses.  Because we are interested in the relative uncertainty in the depth of the transit between bandpasses, we also fix the mass and radius of the star to their best-fit values.  As a result, the best-fit radii in Table \ref{radius_curves_comparison} reflect the \emph{relative} uncertainty in the measurement of the radius in a particular bandpass, and not the total uncertainty in the radius measurement.  This decreased the variations in $R_P$ between bandpasses significantly, to an average of $\pm 0.003$~$R_{\rm Jup}$ or 210~km, with the largest shifts in the $293-347$~nm and $922-1019$~nm bandpasses.  

As we noted earlier, some of the residuals in the data come from the discrepancy between our model limb-darkening coefficients and the observed limb-darkening of the star.  In order to check the effect that the limb-darkening coefficients might have on our results, we repeated our fits using the best-fit quadratic limb-darkening coefficients from the Markov Chain Monte Carlo fit described in \S\ref{transit_curve_fit}.  We found that although the change in radius for a particular bandpass was not always the same as it was for the fits using model limb darkening coefficients, the overall level of variation between bandpasses was virtually identical.

Although we expect absorption lines to produce some differences between bandpasses, a change of 0.003~$R_{\rm Jup}$ in the effective radius of the planet would correspond to an average change of $0.07\%$ in the relative depth of the transit over the entire $50-100$~nm wide bandpass.  By comparison, the sodium absorption line in HD~209458b's atmosphere at 589~nm discovered by \citet{char02} only produces a change of $0.02\%$ in the relative depth of the transit over a range of 1 nm in wavelength space.  Although molecular absorption bands could produce a significant change in the depth of a transit over a wider range in wavelength space, the lines would have to be more than three times stronger than the sodium line discovered by \citet{char02} and cover most of the bandpass to explain the observed variations.  Based on model transmission spectra for HD~209458b \citep{seag00,brown01b,hubb01}, there are only a few metal absorption lines predicted in most of our bandpasses, and the only absorption bands are from water absorption and are located in the longest-wavelength bandpasses of our data.  If these models are accurate, absorption at the level of the variations we measure is unlikely.  

We can also perform a more general estimate of the potential strength of absorption lines.  Using our new value for the radius of the planet and the mass given by \citet{laugh05b} (appropriately scaled to reflect our smaller value for the stellar mass), and the brightness temperature at 24~$\mu$m given by \citet{dem05}, we calculate a scale height of 450~km, or 0.006~$R_{\rm Jup}$.  If the opacity in absorption lines was effectively one over a range of one scale height above the average radius we measure, we would expect to see an increase of 0.006~$\rm R_{\rm Jup}$ in the best-fit radius in the region of the absorption line.  This is twice the average variation we observe.  This means that it is possible, if unlikely given what we already know from the detection of sodium absorption by \citet{char02} and detection limits given by \citet{nar05} in this wavelength range, that this level of variation could be the result of absorption in the atmosphere of HD~209458b.  However, it is entirely possible that smaller variations, on the level of the Na absorption detected by \citet{char02}, might be detectible in the data.  We plan to present the results of a thorough search for spectroscopic transmission features in an upcoming paper.

\subsection{Orbital Ephemeris}\label{timing_results}

We derive a period of $3.52474859\pm0.00000038$ days ($\pm0.033$~s) for the planet and initial transit epoch $T_C=2452826.628521\pm0.000087$~HJD ($\pm7.5$~s).  This period is consistent with the period from \citet{brown01}, $3.52480\pm0.00004$.  We compare our $T_C$ with the results of \citet{brown01}, repeating our fit using the first transit in our dataset as our initial transit time, and find $T_C=2451659.936739\pm0.000087$~HJD.  This differs from the \citet{brown01} value by $1.1\times10^{-5}$ days or 1~s, and was also well within the published error of $10^{-4}$ days (see Table \ref{ephemeris}).  

Our period and time of center of transit are entirely consistent with the values from \citet{brown01}, which use \emph{HST} STIS data alone.  However, our period differs from the period given by \citet{witt05} by 26~s, or 17$\sigma$.  When we use the period and $T_C$ given in \citet{witt05} to calculate the predicted times for the UT 2003 July 5 transit, the value differs by 118 s or 10$\sigma$ from our best-fit time.  We note that our data is more recent and has better time coverage within each transit than the \emph{HST} Fine Guidance Sensor data from 2001 and 2002 that dominates the period measurement by \citet{witt05}.

\begin{figure}
\epsscale{1.0}
\plotone{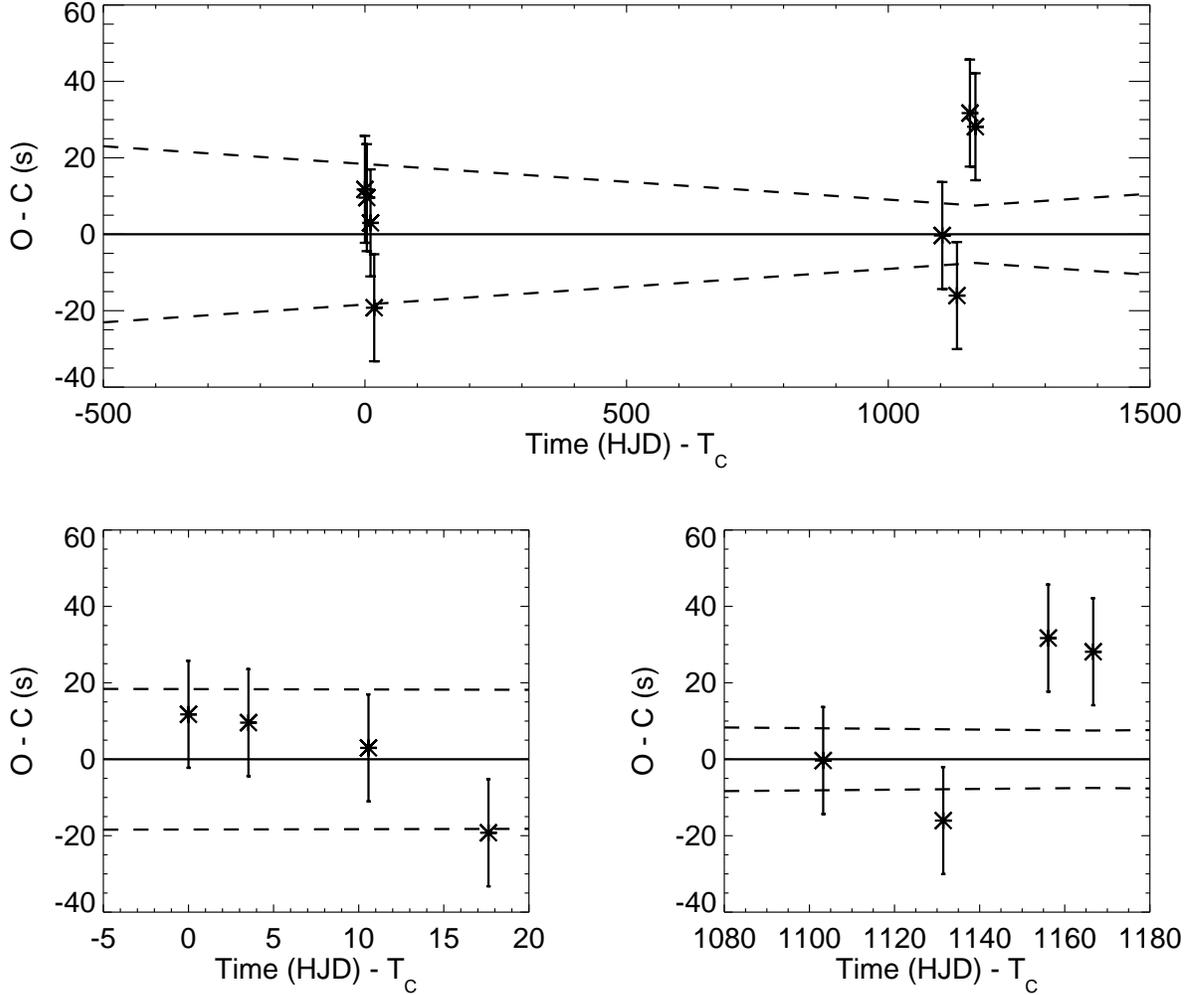}
\caption{These are the O-C residuals for all eight transits using the period and $T_C$ derived in this work.  The errors shown are based on an extrapolation of the systematic spacecraft orbit-to-orbit variations in the data; see \S\ref{timing_results} for a complete description.  The dashed lines are calculated from the uncertainties in the measurements of $P$ and $T_C$. The two lower plots show the relevant regions of the upper plot in more detail.\label{residuals}}
\end{figure}

Although the period for the planet is consistent with previous results by \citet{brown01}, we initially found significant discrepancies between the predicted and measured transit times for individual transits (see Table \ref{hjd_table}).  We derive the errors for our timing measurements using a bootstrap Monte Carlo method, in which we randomly drew measurements with replacement from the data set of a given transit to simulate a data set with the same number of points and noise properties as the original, we then fit for the time of center of transit.  We repeat this process until we obtain an approximately Gaussian distribution of best-fit transit times, and take the standard deviation of this distribution as the error on our timing measurement.  Using this method, we find errors ranging from $3-6$~s.  When we calculate a new $P$ and $T_C$ using these errors, we find that our best-fit transit times differ from their predicted values by as much as 6$\sigma$.  However, as Figure \ref{residuals} demonstrates, there are still some residual trends in the data that are not taken into account by this method.  

We examined the flux residuals from the transit light curves and concluded that there were residual trends in the data that might alter the measured transit times.  Although some of these trends are the result of our choice of limb-darkening coefficients, as discussed in \S\ref{transit_curve_fit}, there are also residuals from our procedure to correct for the instrument variations (described in \S\ref{obs_red}) that might affect the results.  Because systematic shifts over an entire spacecraft orbit will have the largest effect on the transit times, we measured the typical offset of each orbit in the residuals plotted in Figure \ref{transit_residuals}.  We found that the typical offset for an individual orbit was 0.0001 in relative flux.  In order to quantify the effect that this kind of orbit-to-orbit variations in flux might have on the timing measurements, we created artificial data sets for each of the eight transits and shifted the points in each individual spacecraft orbit by a Gaussian random variable with a standard deviation of 0.0001 in relative flux.  For this level of variation, we calculated a standard deviation of $\pm14$s in the best-fit transit times.  Adopting this value as the error on our timing measurements, we find that all shifts are within 2$\sigma$ of their predicted values (see Figure \ref{residuals}).  The errors for $P$ and $T_C$ are derived assuming a constant error of $\pm14$s on each individual transit time.  
				   
\section{Conclusions}

We used multiple-bandpass photometry, the stellar mass-radius relation from \citet{cody02}, and theoretical models for limb darkening to significantly improve the estimates of the radius and orbital inclination of HD~209458b.  We find that the radius of HD~209458b is $1.320\pm0.025$~$R_{\rm Jup}$, a factor of two more precise than previous measurements.  Using the mass of the planet given by \citet{laugh05b} and scaling it appropriately to account for our smaller value for the stellar mass, we find a density for the planet of $0.26\pm0.04$~$\rho_{\rm Jup}$ or $0.345\pm0.05$~g/cm$^3$.  The planet's inclination is $86{\fdg}929\pm0{\fdg}010$, a factor of three more precise than previous measurements.  

We also improve upon previous measurements of the planet's orbital period $P$ and time of center of transit $T_C$.  Using a combination of STIS data spanning four transits in 2000 and four transits in 2003, we find that $P=3.52474859\pm0.00000038$~days ($\pm0.033$~s) and $T_C=2452826.628514\pm0.000087$~HJD ($\pm7.5$~s).  Although the quoted uncertainties in the period for HD~209458b derived by \citet{witt05} are smaller, our data has better signal-to-noise and time coverage than the \emph{Hubble Space Telescope} FGS data which dominates their measurement.  Additionally, the large ($10\sigma$) discrepancy between the transit times predicted by \citet{witt05} and the transit times we measure for the 2003 data indicate that there may be systematics that are not taken into account in the error given by \citet{witt05}.   

When we compare the observed timing of individual transits to the predicted values, we see no discrepancies larger than $2\sigma$.  Our flux residuals and the errors on our timing measurements are modestly larger than the errors quoted by \citet{brown01}, but we extend the time frame of these results by three years and double the number of transits observed at a comparable level of precision.  Using the equations from \citet{hol05}, and setting a limit of 42~s or $3\sigma$ on potential timing variations, we are able to place a limit of 4~$M_{Jup}$ on the mass of a potential second planet in this system, where we have assumed either a relatively short period and circular orbit, or a longer period (up to 100~days) and larger eccentricity (up to $e=0.7$).  We note that this limit is consistent with the smaller limit of 0.3~$M_{Jup}$ that \citet{laugh05b} derive from radial velocity data for this system.  

Given the success of this method, and the need for accurate characterization of extrasolar planet radii, it would be useful to apply this method to other transiting hot Jupiters.  Despite the failure of STIS, recent results by T. Brown (personal communication, 2005) have demonstrated that spectra with a cadence and signal-to-noise comparable to that of STIS can be obtained using the grism spectrometer on \emph{HST}'s Advanced Camera for Surveys.  This opens up the possibility of similar multiple-bandpass studies for other transiting planets, an avenue that we plan to pursue in the future.  There are currently five known extrasolar planets around stars bright enough for this kind of study: HD~209458b, TrES-1 \citep{alon04,soz04}, HD~149026b \citep{sato05,char06}, HD~189733b \citep{bou05,bak06}, XO-1b \citep{mcc06,hol06}, and it is likely that several more will be uncovered in the coming years by a combination of wide-field transit surveys and quick-look radial velocity surveys.

\acknowledgments
We are grateful to R. Kurucz for his assistance in calculating the limb-darkening for HD~209458.  Support for program number HST-GO-09447 was provided by NASA through a grant from the Space Telescope Science Institute, which is operated by AURA under NASA contract NAS5-26555.  H.K. was supported by a National Science Foundation Graduate Research Fellowship.

\begin{deluxetable}{lrrrrcrrrrr}
\tabletypesize{\scriptsize}
\tablecaption{Photometry of HD~209458\label{data_table}}
\tablewidth{0pt}
\tablehead{
\colhead{$\lambda_{center}$ (nm)} & \colhead{HJD} & \colhead{Relative Flux} & \colhead{Uncertainty}
}
\startdata
320.1 &   2452763.055640 &   1.000531 &   0.000609\\
320.1 &   2452763.056126 &   1.000054 &   0.000609\\
320.1 &   2452763.056612 &   1.000691 &   0.000609\\
320.1 &   2452763.057098 &   1.000254 &   0.000609\\
320.1 &   2452763.057585 &   1.000361 &   0.000609\\
\enddata
\tablecomments{The quoted uncertainties are the standard deviation of the out of transit data in each bandpass.  We intend for this Table to appear in entirety in the electronic version of the Astrophysical Journal.  A portion is shown here to illustrate its format.  The data are also available in digital form from the authors upon request.}
\end{deluxetable}

\begin{deluxetable}{lrrrrcrrrrr}
\tabletypesize{\scriptsize}
\tablecaption{Nonlinear limb-darkening coefficients from model\label{limb_dk_coeff}}
\tablewidth{0pt}
\tablehead{
\colhead{$\lambda$ (nm)} & \colhead{c$_1$} & \colhead{c$_2$} & \colhead{c$_3$} & \colhead{c$_4$} 
}
\startdata
$293-347$  & $-0.1015$ & 0.5547 & 0.6096 & $-0.2814$ \\
$348-402$ & 0.0284 & 0.4248 & 0.6646 & $-0.3450$ \\
$403-457$ & 0.2022 & 0.1101 & 0.9690 & $-0.4801$ \\
$458-512$ & 0.3765 & $-0.0119$ & 0.8863 & $-0.4504$ \\
$512-567$ & 0.4957 & $-0.2057$ & 0.9157 & $-0.4340$ \\
$532-629$ & 0.5566 & $-0.2972$ & 0.9190 & $-0.4288$ \\
$629-726$ & 0.6239 & $-0.4176$ & 0.8889 & $-0.4026$ \\
$727-824$ & 0.6495 & $-0.4916$ & 0.8722 & $-0.3844$ \\
$825-922$ & 0.6623 & $-0.5338$ & 0.8411 & $-0.3661$ \\
$922-1019$ & 0.6535 & $-0.5323$ & 0.8142 & $-0.3566$ \\
\enddata
\end{deluxetable}

\begin{deluxetable}{lrrrrcrrrrr}
\tabletypesize{\scriptsize}
\tablecaption{Comparison of limb-to-central intensity \label{solar_vs_model}}
\tablewidth{0pt}
\tablehead{
\colhead{$\lambda$ (nm)} & \colhead{Our model} & \colhead{Best Fit} & \colhead{Solar} 
}
\startdata
$293-347$  & 0.219 & 0.122 & 0.066 \\
$348-402$ & 0.227 & 0.172 & 0.083 \\
$403-457$ & 0.199 & 0.226 & 0.115 \\
$458-512$ & 0.199 & 0.314 & 0.152 \\
$512-567$ & 0.228 & 0.357 & 0.188 \\
$532-629$ & 0.250 & 0.317 & 0.216 \\
$629-726$ & 0.307 & 0.468 & 0.272 \\
$727-824$ & 0.354 & 0.521 & 0.317 \\
$825-922$ & 0.397 & 0.524 & 0.363 \\
$922-1019$ & 0.421 & 0.479 & 0.393 \\
\enddata
\tablecomments{The second column lists the limb-to-central intensity ratio for the four-parameter non-linear limb-darkening coefficients that we use in our fits, calculated from a model for HD~209458.  The third column lists the intensity ratio for the quadratic limb-darkening coefficients that we obtain by fitting the data directly, and the fourth column lists the intensity ratio for theoretical four-parameter non-linear limb-darkening coefficients for the Sun, calculated by \citet{clar00}.}
\end{deluxetable} 

\begin{deluxetable}{lllrlllrrr}
\tabletypesize{\scriptsize}
\tablecaption{Comparison between best-fit values and results from previous works\label{best_fit_coeff}}
\tablewidth{0pt}
\tablehead{
\colhead{Study} & \colhead{R$_P$ (R$_{\rm Jup}$)} & \colhead{Inclination (\degr)} & \colhead{M$_{\star}$ (M$_{\sun}$)} & \colhead{R$_{\star}$ (R$_{\sun}$)}
}
\startdata
\citet{witt05}\tablenotemark{a} & $1.35\pm0.07$\phantom{.} & 86.668\phantom{00000} & $1.09\pm0.09$\phantom{.} & $1.15\pm0.06$\\
\citet{winn205}\tablenotemark{b} & $1.35\pm0.06$\phantom{.} & $86.55\pm0.03$ & $1.06\pm0.13$\phantom{.} & $1.15^{+0.05}_{-0.06}$\phantom{0.}\\
This Work\tablenotemark{a} & $1.320^{+0.024}_{-0.025}$ & $86.929^{+0.009}_{-0.010}$ & $1.101^{+0.066}_{-0.062}$ & $1.125^{+0.020}_{-0.023}$\\
\enddata
\tablenotetext{a}{Used stellar mass-radius relation from \citet{cody02}}
\tablenotetext{b}{Assumed value for the stellar mass from \citet{cody02}}
\phantom{0}
\end{deluxetable}

\begin{deluxetable}{lrrrrcrrrr}
\tabletypesize{\scriptsize}
\tablecaption{Best-fit Planet Radius in Individual Bandpasses\label{radius_curves_comparison}}
\tablewidth{0pt}
\tablehead{
\colhead{$\lambda$ (nm)} & \colhead{Radius ($R_{\rm Jup}$)} 
}
\startdata
$293-347$  & $1.3263\pm0.0018$ \\ 
$348-402$ & $1.3254\pm0.0010$ \\ 
$403-457$ & $1.3200\pm0.0006$ \\ 
$458-512$ & $1.3179\pm0.0006$ \\ 
$512-567$ & $1.3177\pm0.0010$ \\ 
$532-629$ & $1.3246\pm0.0006$ \\ 
$629-726$ & $1.3176\pm0.0005$ \\ 
$727-824$ & $1.3158\pm0.0006$ \\ 
$825-922$ & $1.3200\pm0.0006$ \\ 
$922-1019$ & $1.3268\pm0.0013$ \\ 
\enddata
\tablecomments{The inclination, stellar mass, and stellar radius were set to their best-fit values from Table \ref{best_fit_coeff} for these fits.}
\end{deluxetable}

\begin{deluxetable}{lrrrrcrrrr}
\tabletypesize{\scriptsize}
\tablecaption{Best-fit transit times\label{hjd_table}}
\tablewidth{0pt}
\tablehead{
\colhead{UT Date} & \colhead{$N_T$} & \colhead{$T_C$ (HJD)} & \colhead{$\sigma_{\textrm{HJD}}$} & \colhead{$(O-C)$} & \colhead{$\frac{\textrm{(O-C)}}{\sigma_{\textrm{HJD}}}$}
}
\startdata
2000 April 25 & $-331$ & 2451659.936875 & $\pm0.000162$ & 0.000136 & 0.84 \\
2000 April 28 & $-330$ & 2451663.461599 & $\pm0.000162$ & 0.000111 & 0.68\\
2000 May 5 & $-328$ & 2451670.511019 & $\pm0.000162$ & 0.000034 & 0.21\\
2000 May 12 & $-326$ & 2451677.560259 & $\pm0.000162$ & $-0.000223$ & $-1.37$ \phantom{0}\\
2003 May 3 & $-18$ & 2452763.183042 & $\pm0.000162$ & $-0.000004$ & $-0.02$ \phantom{0}\\
2003 May 31 & $-10$ & 2452791.380849 & $\pm0.000162$ & $-0.000186$ & $-1.15$ \phantom{0}\\
2003 June 25 & $-3$ & 2452816.054642 & $\pm0.000162$ & 0.000367 & 2.26\\
2003 July 5 & 0 & 2452826.628846 & $\pm0.000162$ & 0.000326 & 2.01\\
\enddata
\tablecomments{These are the best-fit locations for the centers of the eight eclipses examined in this work.  The errors are based on our estimate of the size of residual trends in the data, which we believe to be the largest source of error.  We also give the number of elapsed transits and O-C residuals for each eclipse.}
\end{deluxetable}

\begin{deluxetable}{lrrrrcrrrrr}
\tabletypesize{\scriptsize}
\tablecaption{Ephemeris from this work and previous works.\label{ephemeris}}
\tablewidth{0pt}
\tablehead{
\colhead{Study} & \colhead{Period (days)} & \colhead{$T_C$ (HJD)}
}
\startdata
\citet{brown01} & 3.52480\phantom{000}$\pm0.00004$\phantom{000} & 2451659.93675\phantom{0}$\pm0.0001$\phantom{00}\\
\citet{witt05} & 3.52474554$\pm0.00000018$ & 2452854.82545\phantom{0}$\pm0.000135$\\
This Work & 3.52474859$\pm0.00000038$ & 2452826.628521$\pm0.000087$\\
\enddata
\end{deluxetable}

\end{document}